\documentclass[twocolumn, times, preprint]{aastex6}

\usepackage{amsmath, bm,soul}

\newcommand{\ubar}{$\bar u$ }
\newcommand{\vbar}{$\bar v$ }
\newcommand{\phiby}[1]{$\phi=\frac{\pi}{#1}$}

\begin{document}


\title{The effect of magnetic field on mean flow generation by rotating two-dimensional convection}


\author{Laura K. Currie}
\affil{Department of Physics and Astronomy, University of Exeter, Stocker Road, EX4 4QL Exeter, UK; lcurrie@astro.ex.ac.uk}
\altaffiltext{}{lcurrie@astro.ex.ac.uk}





\begin{abstract}
Motivated by the significant interaction of convection, rotation and magnetic field in many astrophysical objects, we investigate the interplay between large-scale flows driven by rotating convection and an imposed magnetic field. We utilise a simple model in two dimensions comprised of a plane layer that is rotating about an axis inclined to gravity. It is known that this setup can result in strong mean flows; we numerically examine the effect of an imposed horizontal magnetic field on such flows. We show that increasing the field strength in general suppresses the time-dependent mean flows, but in some cases it organises them leading to stronger time-averaged flows. Further, we discuss the effect of the field on the correlations responsible for driving the flows and the competition between Reynolds and Maxwell stresses. A change in behaviour is observed when the (fluid and magnetic) Prandtl numbers are decreased. In the smaller Prandtl number regime, it is shown that significant mean flows can persist even when the quenching of the overall flow velocity by the field is relatively strong.
\end{abstract}




\section{Introduction}
Many astrophysical flows are turbulent and contain systematic large-scale (mean) flows that reside alongside smaller-scale turbulent eddies. Well-known examples of such large-scale flows are the zonal jets evident at the surface of the gas giants (e.g., \citet{Porcoetal2003, VasavadaShowman2005}) and the strong zonal and meridional flows in the interior of the Sun which are understood as the observed differential rotation and meridional circulations (\cite{Schou1998}). 
It is widely accepted that the dynamics of these flows is often further complicated by the presence of a magnetic field.

Techniques such as spectropolarimetry, asteroseismology, and photometric monitoring have allowed observers to probe differential rotation in stars and its interaction with magnetism. 
For example, \cite{Reinholdetal2013} use precision photometry to obtain surface differential rotation that varies with spectral type in a large sample of Kepler stars. Meanwhile, spectropolarimetric measurements of surface differential rotation in very low-mass stars (see e.g., \cite{Donati2003, DonatiLandstreet2009}) often suggest that such objects rotate nearly as solid bodies, with quenching of zonal flows by magnetic fields being a possible cause. 
Asteroseismology of red giants (e.g., \cite{Becketal2012, Deheuvelsetal2012}) has also revealed strong internal differential rotation, with the cores of some objects rotating multiple times faster than their envelopes.  Separately, global simulations of turbulent convection under the influence of rotation have suggested multiple regimes of zonal flows are possible, particularly in the presence of magnetic fields: in some cases angular velocity contrasts persist even in the presence of relatively strong magnetic fields while in others the large-scale flows are largely eliminated.  What delineates these regimes, and determines the amplitude and direction of the large-scale flows, is still not known (see e.g., \cite{Brunetal2005, Kapylaetal2011, Gastineetal2014, Karaketal2015}). 
It is clear then, that in addition to understanding the interaction between the smaller-scale turbulence and the large-scale flows, it is important to determine the role of a magnetic field in such a system. Such interactions are complicated, and although much progress has been made (as discussed briefly below) there is still much that is not understood.

In order to evaluate the effects of a magnetic field on mean flows, here we consider flows driven by convection resulting from an imposed temperature gradient. Such is the complexity of modelling the interactions of convection, rotation and magnetic fields, a starting point has often been the hydrodynamic problem whereby the effects of magnetic fields are neglected. Mean flow generation in hydrodynamic models, in both local and global geometries, has received significant attention in the literature: (see e.g., \cite{HS1983, HS1986, HS1987, JulienKnobloch1998, SaitoIshioka2011} (local) and \cite{Mieschetal2000, Elliottetal2000, Christensen2001, Christensen2002, BrunToomre2002, Browningetal2004, GastineWicht2012, Gastineetal2013} (global).) The mechanism for the generation of such large-scale flows is dependent upon the system of study, and in particular the geometry. In rotating spherical shell models zonal flow is thought to be driven by Reynolds stresses resulting from the curvature of the boundaries (\cite{Busse1983}); although density variations may also provide a source of vorticity that can help to sustain mean flows (e.g., \cite{Evonuk2008, Gastineetal2014a, VerhoevenStellmach2014}). In order to capture some important geometrical effects of a spherical body but whilst maintaining a relative simplicity, \cite{Busse1970} introduced an annulus model. This setup has since been implemented in models of the zonal flow on Jupiter. For example, \cite{Jonesetal2003} used a two-dimensional (2d), rotating annulus setup which allowed for more realistic jet solutions to be found when boundary friction was included. \cite{RotvigJones2006} explored this annulus model further and identified a bursting mechanism that occurs in the convection in some cases.
In a local Cartesian geometry, \cite{HS1983, HS1986, HS1987} studied the flows generated when the rotation vector was oblique to gravity in a number of different models. The plane layer geometry, when the axis of rotation is allowed to vary from the direction of gravity, is often used to approximate a local region of fluid located at different latitudes of a spherical body; this introduces an asymmetry into the system. This asymmetry is enough to drive significant mean flows within the system. As an additional mechanism, \cite{Currie2014} examined mean flow generation when a thermal wind (driven by the presence of horizontal temperature gradients) was present. \cite{CT2016} extended the work of \cite{HS1983} to include the effects of a background density stratification. Furthermore, strong shear flows can be driven in 2d models of Rayleigh-B\'enard convection that employ horizontally periodic, and vertically stress-free, boundary conditions (see \cite{Goluskinetal2014}).

The effect of a magnetic field on convection has received less attention. Whilst there is a vast body of literature on magnetic field generation in astrophysical objects through dynamo action (e.g., \cite{Moffatt1978}, \cite{Parker1979}, \cite{Tobias2002}, \cite{BrandenburgSubramanian(2005)}, \cite{Jones2011}), there is less relating to the effect of the magnetic field on the convection and in particular, the effect of an imposed field on the driving and maintaining of mean flows seen in hydrodynamic systems. In this paper we impose a horizontal magnetic field and assess its effect on the system; we do not address the question of how the field got there. The difficulties in solving the full dynamo problem make magnetoconvection in an imposed magnetic field an important tool for studying the basic principles which influence the interactions between convection and magnetic fields.

Early studies of magnetoconvection include the linear analyses of \cite{Chandrasekhar1961}, \cite{Eltayeb1972, Eltayeb1975}. Eltayeb derived a bound for which rotational effects dominate over magnetic effects in a plane layer system with rotation and magnetic field both in the horizontal direction. \citet{Arter} studied 2d nonlinear convection in an imposed horizontal magnetic field in a plane layer but without rotation. He found that, in general, stronger horizontal magnetic fields resulted in time-dependent convection and, as the thermal driving was increased, the oscillations grew in amplitude until the flow direction reversed. An analogous problem to the one studied by \cite{Arter} involves convection in an imposed vertical field; this topic was comprehensively reviewed by \citet{PW}. \citet{Arter} highlighted that an imposed horizontal magnetic field results in convective motions which are significantly different to those in a vertical field, since in the latter case, flux can separate out from the flow.

Later studies of magnetoconvection (often motivated by the need to understand convection in the Sun) looked to include compressibility effects. For example, \cite{LantzSudan1995} used numerical simulations to solve anelastic equations in an imposed horizontal magnetic field but without the effects of rotation. \cite{HurlburtToomre1988} considered nonlinear fully compressible convection in an imposed vertical field, again without rotation. They found the convection sweeps the initially vertical field into concentrated flux sheets and, for strong enough imposed fields, the Lorentz force can suppress the flows.
In global calculations, a main focus of study has been dynamo theory and so much of the existing literature does not examine convection in an imposed field (e.g., \cite{Brunetal2005}, \cite{Browning2008}, \cite{Brownetal2011}, \cite{FanFang2014}). However, the early global simulations of \cite{OlsonGlatzmaier1995} did consider convection in an imposed toroidal field under the Boussinesq approximation.

In this paper we focus on mean flows driven by convection in a 2d rotating plane layer in which rotation is oblique to gravity and the layer is permeated by an imposed horizontal magnetic field. It is believed that large-scale magnetic fields that emerge in sunspots or in active regions on the Sun originate near the base of the convection zone where the field is mostly azimuthal (see e.g., \cite{GallowayWeiss1981}) and therefore we chose to impose a horizontal (and not, as is often done, a vertical) magnetic field.
As described above, the tilted plane layer geometry has been shown to be capable of sustaining mean flows, and here we assess the impact of an imposed horizontal field on such flows. For simplicity, we will consider Boussinesq fluids only. The simplicity of this setup, compared to other more complicated (and realistic) geometries, allows us to study some parameter regimes more easily and to identify key physical interactions between magnetic field and mean flows driven by rotating convection. 

The main aim of this article then is to elucidate the role of magnetic field in modifying the mean flows driven self-consistently by convection. In section \ref{model}, the model and governing equations are presented. In sections \ref{results1} and \ref{results2}, we present results from numerical simulations, highlighting the role of magnetic field in different parameter regimes. Finally, in section \ref{conc} we offer conclusions and a discussion of the results and how they may be developed in future.

\section{Model setup and equations}\label{model}

\begin{figure}[hbt]
\begin{center}
\includegraphics[scale=1]{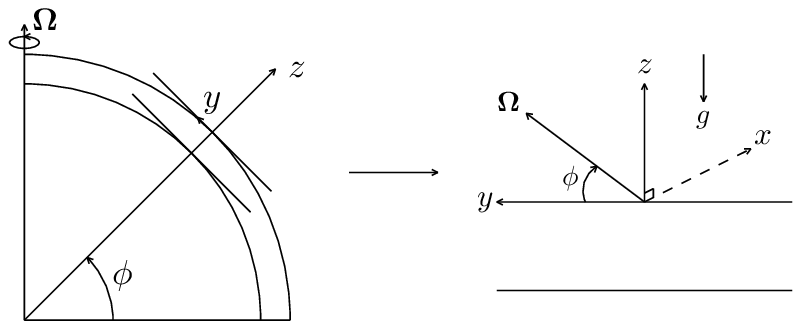}
\caption{A schematic of the model setup. We employ a Cartesian plane layer centred about a latitude $\phi$ of a spherical body that is rotating with velocity $\bm{\Omega}=(0,\Omega\cos\phi,\Omega\sin\phi)$. $x$ is directed eastwards, $y$ is directed northwards and $z$ points upwards.}\label{config}
\end{center}
\end{figure}

We consider a local Cartesian layer of Boussinesq fluid rotating about an axis that is
oblique to gravity which acts downwards (in the negative
$z$-direction). The rotation vector lies in the $y$-$z$ plane and is given by
$\bm{\Omega}=(0,\Omega\cos\phi,\Omega\sin\phi)$, where $\phi$ is the angle of inclination of
the rotation vector as measured from the horizontal. Therefore, the layer can be considered to be located tangent to a sphere at a latitude $\phi$. In this case, the $z$-axis points upwards, the $x$-axis eastwards and the $y$-axis northwards (see figure \ref{config}). In addition, we impose a vertical temperature gradient to drive convection and a horizontal magnetic field, appropriate for modelling stellar interiors. The imposed temperature and magnetic fields are given respectively as
$T_{BS}=T_o-\Delta T z$ and $\mathbf{B}_{BS}=B_o(0,1,0)$, where $\Delta T$ is the imposed temperature difference across the layer, $T_o$ is the value of $T$ on the bottom boundary and $B_o$ is the magnitude of the imposed magnetic field. Note that the imposed field is purely in the $y$-direction.

We non-dimensionalise quantities using the layer depth $d$ as the unit of length, the thermal diffusion time $\frac{d^2}{\kappa}$ as the unit of time ($\kappa$ is the thermal diffusivity), temperature with $\Delta T$ and magnetic field with $B_o$. The dimensionless equations that govern the convective motions are then given by the momentum equation, continuity equation, energy equation, induction equation and solenoidal constraint respectively as:
\begin{align}\label{Bousmomeq}\nonumber
\frac{\partial \mathbf u}{\partial t} &+(\mathbf u\cdot \nabla)\mathbf
u=-Pr\nabla p+RaPrT\mathbf{\hat e_z}-TaPr\bm\Omega\times\mathbf
u\\ 
&+Q\zeta Pr[(\nabla \times
\mathbf B)\times(\mathbf B_{BS}+\mathbf B)]+Pr\nabla^2\mathbf u,
\end{align}
\begin{equation}\label{incompeq}
 \nabla\cdot\mathbf u=0,
\end{equation}
\begin{equation}\label{tempeq}
 \frac{\partial T}{\partial t}+(\mathbf u \cdot \nabla)T-w=\nabla^2T,
\end{equation}
\begin{equation}
 \frac{\partial \mathbf B}{\partial t}=\nabla \times [\mathbf u \times (\mathbf B_{BS}+\mathbf B)]+\zeta
\nabla^2\mathbf B,
\end{equation}
\begin{equation}\label{divBeq}
 \nabla\cdot\mathbf B=0,
\end{equation}
where $\mathbf{u}=(u,v,w)$, $p$, $T$ and $\mathbf{B}=(B_x,B_y,B_z)$ denote the fluid velocity, pressure, temperature and magnetic field respectively. Further, 
we have introduced the following dimensionless numbers:
\begin{align}\nonumber
 &Ra=\frac{g \alpha d^3\Delta T}{\kappa\nu}, \quad
Pr=\frac{\nu}{\kappa}, \quad Ta=\frac{4\Omega^2d^4}{\nu^2}, \\
&Q=\frac{B_0^2d^2}{\mu_0\rho_0\nu\eta}, \quad \zeta=\frac{\eta}{\kappa}.
\end{align}
In these expressions $\alpha$ is taken to mean the coefficient of volumetric expansion, $g$ the acceleration due to gravity, $\nu$ the kinematic viscosity of the fluid, $\mu_0$ is the permeability of free space and $\eta$ the magnetic diffusivity. 
$Ra$, $Pr$, $Ta$ and $Q$ are the usual Rayleigh, (fluid) Prandtl, Taylor and Chandrasekhar
numbers. $\zeta$ is the ratio of magnetic diffusivity to thermal diffusivity; note the inverse of this quantity is sometimes referred to as the Roberts number. We note
here that other commonly used dimensionless numbers can be obtained from $Ra$, $Pr$, $Ta$ and $Q$. The magnetic Prandtl, Elsasser and Rossby numbers are given respectively by
\begin{equation}
 Pm=\frac{\nu}{\eta}=\frac{Pr}{\zeta}, \quad
\Lambda = \frac{Q}{Ta^{\frac{1}{2}}}, \quad 
Ro=\left(\frac{Ra}{Ta Pr}\right)^{\frac{1}{2}}.
\end{equation}

We impose impenetrable, stress free, fixed temperature and perfectly conducting boundary conditions on the top and bottom boundaries, i.e., 
\begin{align}
w=\frac{\partial u}{\partial z}=\frac{\partial v}{\partial z}=T=\frac{\partial B_x}{\partial z}=\frac{\partial B_y}{\partial z}=B_z=0 \quad \text {on $z=0,1$}
\end{align}
and we assume all variables to be periodic in the horizontal directions.




To solve the nonlinear system of equations given by (\ref{Bousmomeq})-(\ref{divBeq}), we restrict ourselves to the two-dimensional case where the variables are assumed to be independent of $x$.
This allows us to introduce a streamfunction, $\psi(y,z)$, defined by
\begin{equation}
\bm u = u \bm{\hat x}+\nabla \times \psi(y,z) \bm{\hat x}
=\left(u, \frac{\partial \psi }{\partial z}, -\frac{\partial \psi}{\partial y}\right),
\end{equation}
which automatically satisfies $\nabla \cdot \mathbf{u}=0 $.
In an analogous way, we introduce a flux function $A(y,z)$, defined by 
\begin{equation}
\bm B =B_x \bm{\hat x}+ \nabla \times A(y,z) \bm{\hat x} 
=\left(B_x, \frac{\partial A}{\partial z}, -\frac{\partial A}{\partial y}\right),
\end{equation}
so that $\nabla \cdot \mathbf{B}=0 $.
We then express the equations in terms of $\psi$, $\omega=-\nabla^2\psi$, $u'$, $T'$, $B_x'$, $A$ and $j=-\nabla^2A$ and solve them for these time-dependent variables using a Fourier-Chebyshev pseudospectral method with a a second order, semi-implicit, Crank-Nicolson/Adams-Bashforth time-stepping scheme (see e.g., \cite{Boyd}, \cite{Peyret}). It is then straightforward to obtain $v$ and $w$ from $\psi$ and likewise $B_y$ and $B_z$ from $A$.




To aid our analysis, we split the variables into a mean (horizontally averaged) part and a fluctuating part, where the mean (denoted by overbar) is defined as, for example,
 \begin{equation}
  \bar u(z,t)=\frac{1}{L}\int_0^Lu(y,z,t) \, dy.
 \end{equation}
 The kinetic energy of the velocity perturbations is given by
 \begin{align}\label{KEpert}\nonumber
K&E_{\text{pert}}(t)=\\
 &\frac{1}{2L}\int_0^1\int_0^L(u(y,z,t)^2+v(y,z,t)^2+w(y,z,t)^2) \, dy \, dz
\end{align}
and similarly, we define the magnetic energy in the perturbations by
\begin{align}\label{MEpert}\nonumber
 M&E_{\text{pert}}(t)=\frac{ Q\zeta Pr}{2L} \times\\
&\int_0^1\int_0^L({B_x}(y,z,t)^2+B_y(y,z,
t)^2+B_z(y,z,t)^2) \, dy \, dz.
\end{align}
Note (\ref{KEpert}) and (\ref{MEpert}) contain the energy in both the mean and fluctuating parts but not, in the case of (\ref{MEpert}), the basic state; though this can be added to $ME_{\text{pert}}$ without difficulty.

As this article is largely concerned with mean flow generation, it will be useful to define the size of any mean flows generated. We consider two measures:
 \begin{equation}  \label{one}
 1. \quad \langle \bar \xi \rangle_{rms} = (\{\langle \bar \xi \rangle ^2\})^{\frac{1}{2}} \end{equation} 
  \begin{equation} \label{two}
  2. \quad \bar \xi_{r\langle m \rangle s} = (\langle \{ \bar {\xi} \,^2\}\rangle)^{\frac{1}{2}} \end{equation} 
where $\xi$ is the variable $u$ or $v$, $\{ \cdot \}$ represents an average over the layer depth and $\langle \cdot \rangle$ denotes a time-average.
The first definition (given by equation (\ref{one})) is a measure of the mean of $\bar \xi$; positive and negative contributions to $\bar \xi$ cancel on time averaging and so this measure gives a guide to the magnitude of the time-averaged flow. The second definition (given by equation (\ref{two})) is a measure of the time-dependent $\bar \xi$; since $\bar \xi$ is squared first, positive and negative quantities both contribute. Measures (\ref{one}) and (\ref{two}) taken together give information about the mean and variability of the large-scale flows driven, we therefore consider both measures in our analysis.  For example, a flow that is directed northwards, say, for all time, will have a much larger $\langle \bar \xi \rangle_{rms}$ than a flow of the same magnitude that alternates northwards and southwards but the two flows would have the same  $\bar \xi_{r\langle m \rangle s}$. We consider a flow with similar values of $\langle \bar \xi \rangle_{rms} $ and $ \bar \xi_{r\langle m \rangle s}$ to be systematic (having less variability in time). To quantify how systematic a flow is we define
\begin{equation}\label{sigma}
\sigma_{\bar \xi}=\frac{\langle \bar \xi \rangle_{rms}}{\bar \xi_{r\langle m \rangle s}}.
\end{equation}
Hence, the closer to unity $\sigma_{\bar \xi}$ is, the more systematic the flow.

The size of the mean flow as given by (\ref{one}) or (\ref{two}) can be compared with the size of the flow itself. To enable us to do this we define a typical flow velocity as
\begin{equation}\label{xityp}
|\xi|=(\{\langle \bar \xi^2 \rangle \})^{\frac{1}{2}}.
\end{equation}



\section{Numerical results at moderate Prandtl numbers}\label{results1}
\begin{figure}[hbt]
\begin{center}
\includegraphics[scale=1]{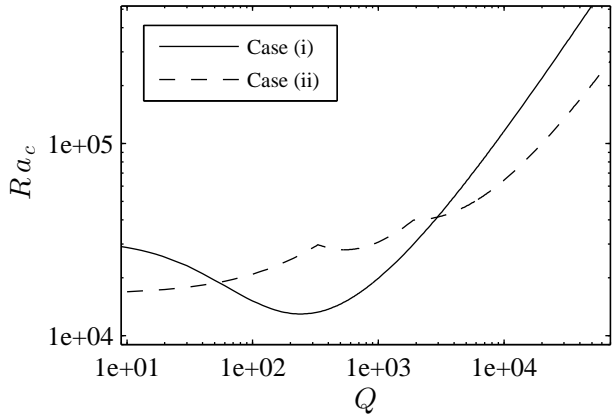}
\caption{Critical Rayleigh number, $Ra_c$, against $Q$ for case (i): $Pr=1$, $\zeta=1.1$, $Ta=10^5$, $\phi=\frac{\pi}{4}$ (solid line) and case (ii): $Pr=0.1$, $\zeta=0.5$, $Ta=5\times10^5$, $\phi=\frac{\pi}{4}$ (dashed line).}\label{Critical}
\end{center}
\end{figure}

For all results presented here we fix the angle of the rotation vector to \phiby4 (representative of mid-latitudes) and fix the aspect ratio of the computational domain by setting $\frac{L}{d}=5,$ where $L$ is the width of the domain. We initially consider $Ta=10^5$, $Pr=1$ and $\zeta=1.1$ (corresponding to $Pm=0.91$), but the effect of changing $Pr$ and $\zeta$ are considered in section \ref{results2}.
To begin, we briefly consider the effect of a horizontal magnetic field on the velocity field of the system.
The effect of the field can be seen by taking a hydrodynamic simulation (equivalent to $Q=0$) and increasing $Q$, thus increasing the strength of the magnetic field. We find, as expected, that as the strength of the field is increased (so that the magnetic energy of the system
is also increased), the kinetic energy decreases. Since the basic state field
lies in the $y$-direction, any attempt by a flow in the $x$-direction to draw out field
lines is opposed by the field. This results in the flow in the $x$-direction being reduced
and hence contributes to the decrease in the kinetic energy we observe. 
It is typical for us to find (at least in the $Pr=1$, $\zeta=1.1$ case) that the solutions at small $Q$ are chaotic but as $Q$ is increased (at fixed $Ra$), the solutions eventually become steady. 
Increasing the size of the imposed magnetic field (i.e., increasing $Q$) has an effect on the critical Rayleigh number ($Ra_c$) (see e.g., \cite{Chandrasekhar1961, RobertsJones2000}). For reference, $Ra_c$ is shown in figure \ref{Critical} as a function of $Q$ for the cases considered in this paper. Note that $Ra_c$ exhibits non-monotonic behaviour at small to moderate $Q$, but for large $Q$, increasing $Q$ also increases $Ra_c$. 

\begin{figure*}
\begin{center}
\includegraphics[scale=1]{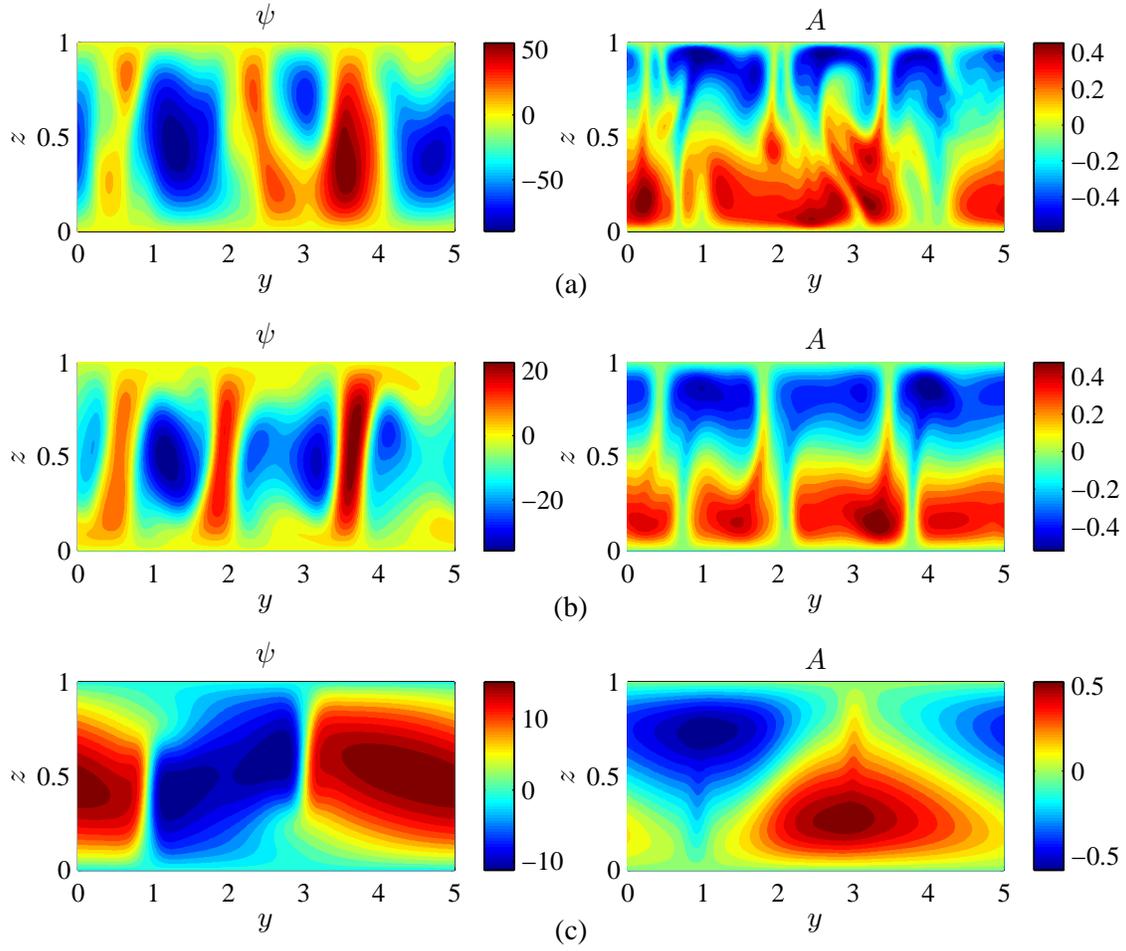}
\caption{Contours of $\psi(y,z)$ (left-hand column) and $A(y,z)$ (right-hand column) at a
snapshot in time for $Pr=1$, $\zeta=1.1$, $Ra=5\times10^5$, $Ta=10^5$, $\phi=\frac{\pi}{4}$ and (a) $Q=100$, (b)
$Q=1500$ and (c) $Q=10000$. (a) and (b) correspond to chaotic solutions and (c) is a
steady solution.}\label{magsol}
\end{center}
\end{figure*}

To visualise the flow and the magnetic field as $Q$ is increased, we have plotted contours
of the streamfunction $\psi(y,z)$ and the flux function $A(y,z)$, at a snapshot in time,
for three different values of $Q$ (see figure \ref{magsol}). Whilst not displayed here, the temperature field has features that track the velocity field well; upflows transport hotter fluid and downflows cooler fluid. In (a), $Q=100$, and
therefore the solution only differs slightly from the solution in the purely hydrodynamic
case and is chaotic; in (b), $Q=1500$, and the solution is still chaotic but this solution occurs just before the solutions go steady. In (c), $Q=10000$
and these solutions are now steady. In these cases, $Ro$ is moderate, which means there is little evidence of the tilting of convection cells that is present when rotation plays more of a role. Nonetheless we see that as $Q$ is increased, the field organises, and reduces the magnitude of, the flow, so that it eventually becomes steady. In doing so,
the length scale of the solution increases from being such that three pairs of negative
and positive cells fit in the box at $Q=100$ to just one pair fitting in the box by
$Q=10000$. A linear calculation of the wavenumber, $l$, of the fastest growing mode at $Ra=5\times10^5$ highlights the effect of the nonlinear terms (note, $l$ is not necessarily the same as the critical wavenumber since convection onsets at smaller $Ra$). For the
cases in figure \ref{magsol} such a calculation gives: $l=6$ for case (a), $l=5$ for case (b) and $l=3$ for
case (c), all representing structure on smaller scales than is actually realised in the fully nonlinear calculation. In other words, the nonlinear terms have acted to increase the length scale of the solutions we observe.

\subsection{Effect of the magnetic field on the mean flows}\label{mf}
\begin{figure*}
\begin{center}
\includegraphics[scale=1]{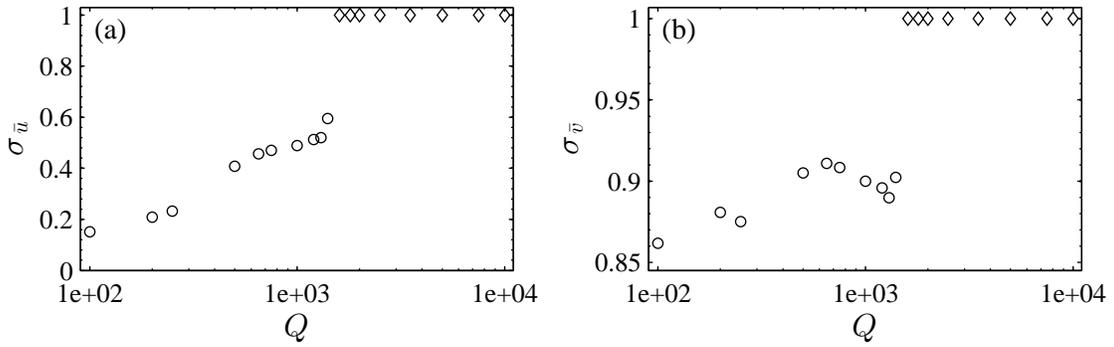}
\caption{(a) $\sigma_{\bar u}$ and (b) $\sigma_{\bar v}$ against $Q$ for $Pr=1$, $\zeta=1.1$, $Ra=5\times10^5$, $Ta=10^5$ and  $\phi=\frac{\pi}{4}$. The circles represent chaotic solutions whilst diamonds represent steady solutions. In the chaotic regime $\sigma_{\bar v}$ is up to 5.75 times larger than $\sigma_{\bar u}$.}\label{systematic}
\end{center}
\end{figure*}

As described in the introduction, we wish to examine the effect of a horizontal magnetic field on the mean flows that are driven self-consistently within the system. To begin, we consider how $\sigma_{\bar \xi}$ (defined in (\ref{sigma})) changes with $Q$. Figure \ref{systematic} shows the results for both $\bar u$  and $\bar v$ for $Pr=1$, $\zeta=1.1$, $Ra=5\times10^5$ and $Ta=10^5$ and a range of $Q$. For $\bar u$, as $Q$ (i.e., the strength of the imposed magnetic field) is increased, $\sigma_{\bar u}$ increases in the chaotic regime until the steady regime is reached where  $\sigma_{\bar u}=1$, by definition. We note that in the chaotic regime, as $Q$ is increased, the supercriticality of the flow first increases and then decreases (cf. figure \ref{Critical}, solid line) yet $\sigma_{\bar u}$ shows monotonic behaviour in this regime. In other words, the magnetic field acts to organise the flow so that it becomes more systematic in time. 
For $\bar v$, there is a general upwards trend for  $\sigma_{\bar v}$ with increasing $Q$ in the chaotic regime but all the points in this regime lie within $3.6\%$ of the average value of 0.893 and so we conclude that the field has little influence on how systematic $\bar v$ is here. It is clear that $\bar v$ is more systematic than $\bar u$, since  $\sigma_{\bar v}\geq \sigma_{\bar u}$ for any fixed $Q$; in fact, in the chaotic regime, $\sigma_{\bar v}$ is up to 5.75 times larger than  $\sigma_{\bar u}$. This can perhaps be explained by the fact $\bar v$ is the mean flow component in the plane of the rotation vector and so in some sense has less freedom to fluctuate than $\bar u$.

\begin{figure*}
\begin{center}
\includegraphics[scale=1]{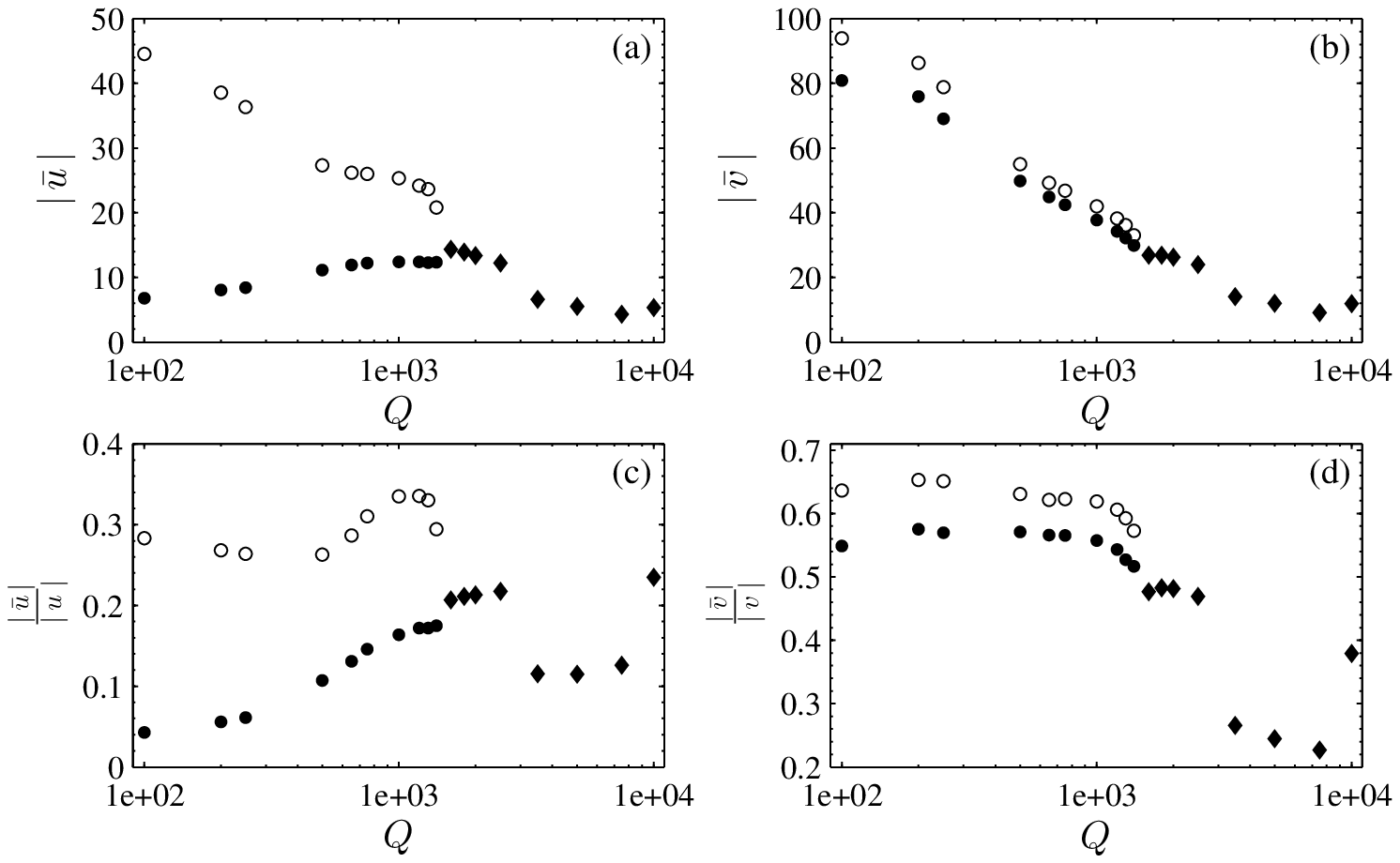}
\caption{Size of $\bar u$ and $\bar v$ for the same parameters as in figure \ref{systematic}. (a) shows the size of $\bar u$ as given by (\ref{one}) ($\langle \bar u \rangle_{rms}$, closed symbols) and (\ref{two}) ($\bar u_{r\langle m \rangle s}$, open symbols). (b) shows the equivalent for $\bar v$. (c) gives the ratio of the size of $\bar u$ to the size of $u$ for the size of $\bar u$ given by (\ref{one}) ($\langle \bar u \rangle_{rms}/|u|$, closed symbols) and by (\ref{two}) ($\bar u_{r\langle m \rangle s}/|u|$, open symbols). (d) shows the equivalent ratios for $\bar v$. The shape of the symbol has the same meaning as in figure \ref{systematic}.}\label{mfvQ}
\end{center}
\end{figure*}

It is informative to consider the effect the horizontal field has on both the strength and direction of the mean flows. Figure \ref{mfvQ} (a) and (b) shows the average size of $\bar u $ and $\bar v$ as calculated by the two measures given by (\ref{one}) and (\ref{two}). We focus on the results in the chaotic regime (given by circles) as we expect flows in reality to be chaotic (rather than steady) and this is the regime in which the most significant mean flows are driven. Both $\bar u_{r\langle m \rangle s}$ and $\bar v_{r\langle m \rangle s}$ decrease as $Q$ is increased indicating that the magnitude of the time-dependent mean flow is, in general, decreased. Owing to $\bar u$ becoming more systematic with increasing field strength, $\langle \bar u \rangle_{rms}$ actually increases with $Q$. In contrast, $\langle \bar v \rangle_{rms}$ decreases as $Q$ is increased, i.e., the field acts to suppress the time-averaged mean flow in this direction.

Since the effect of the field is to quench the velocity field, this effect is likely to contribute to a decrease in $\bar \xi_{r\langle m \rangle s}$ but this then poses the question: does increasing $Q$ decrease the size of $\bar u$ and $\bar v$ just by reducing $u$ and $v$ themselves, or are there other processes affecting the generation of mean flows? To understand this, we consider the ratio of each measure of the strength of the mean flow to a measure of the flow itself, as given by $|\xi|$, defined in equation (\ref{xityp}) (see figure \ref{mfvQ} (c) and (d)). Since changing $Q$ can affect both the magnitude and direction of the mean flows, we consider each measure separately. 
From figure \ref{mfvQ} (c), we do not conclude anything definitive about $\bar u$; since $\langle \bar u \rangle_{rms}$ is increasing with $Q$ as $|u|$ decreases, their ratio must also increase with $Q$ and this is what we see. For $\bar u_{r\langle m \rangle s}/|u|$, the ratio oscillates about the average of approximately 0.3. For $\bar v$ there is a relatively clearer trend in which both ratios maintain a roughly constant value (if not, tend to decrease) with increasing field strength. We therefore suggest that, in this regime, the correlations driving $\bar v$ are suppressed by the field as much as (if not, more severely than) $v$ itself.

\begin{figure*}
\begin{center}
\includegraphics[scale=1]{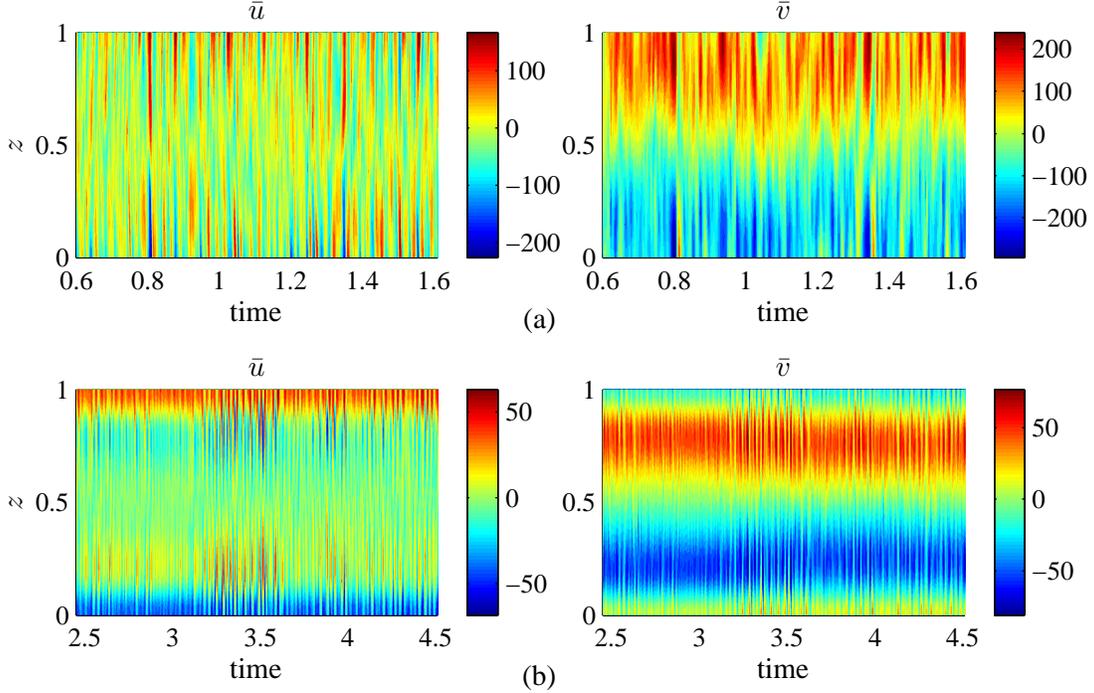}
\caption{\ubar (left) and \vbar (right) as a function of $z$ and $t$ for $Pr=1$, $\zeta=1.1$,
$Ra=5\times10^5$, $Ta=10^5$,  $\phi=\frac{\pi}{4}$ and (a) $Q=100$, (b) $Q=1500$.}\label{tdmf}
\end{center}
\end{figure*}

In addition to studying the time-averaged properties of the mean flows, it is instructive to analyse the
time-dependent mean flows as this can give important information about the nature of
the flows that may not otherwise be captured. For example, figure
\ref{tdmf} shows \ubar and \vbar as a function of $z$ and $t$ for a case when the field
strength is (a) small ($Q=100$), and (b) moderate ($Q=1500$). Both examples are taken from
the chaotic regime of the examples used in figure \ref{magsol}. For small $Q$, case (a),
the mean flows are very similar to those seen in the hydrodynamic system (see e.g., \cite{CT2016}). In particular, \vbar is more systematic
than \ubar and is predominantly positive in the upper half-plane and predominantly
negative in the lower half-plane. In case (b), the magnetic field strength is increased and the nature of \ubar and \vbar has changed. Firstly, let us
consider $\bar v$: whilst there is still a band of positive flow in the upper half-plane
and a band of negative flow in the lower half-plane, the bands do not extend all the way
to the top and bottom boundaries, as they did when $Q=100$ (a). As $Q$ has increased,
boundary layers have formed where the flow has been significantly reduced. The behaviour
that causes this change to occur will be discussed in section \ref{sec:mfeqn}. Secondly,
we also observe a change in the nature of $\bar u$; boundary layers are also formed in
this case, a layer of positive flow at the top boundary and a layer of negative flow at
the bottom boundary. But, in contrast to $\bar v$, the flow is largest in these layers.
Further away from the boundaries, a negative band is evident in the top half of the plane
and a positive band in the lower half of the plane. These bands are more coherent than any
seen in \ubar when $Q=100$; this highlights the fact that increasing $Q$ organises the
flow into having a more systematic nature. 
From (a) and (b) the limits of the colorbar (maximum amplitude of the flow) are decreased as $Q$ is increased, supporting the fact that $\bar \xi_{r\langle m \rangle s}$ is reduced when the imposed magnetic field strength is increased.

\begin{figure}
\begin{center}
\includegraphics[scale=0.95]{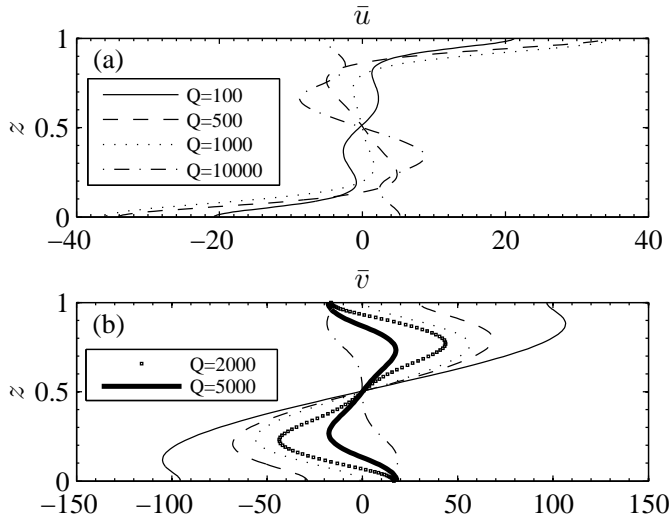}
\caption{$\langle \bar u \rangle$ (a) and $\langle \bar v \rangle$ (b) for $Pr=1$, $\zeta=1.1$, $Ra=5\times10^5$, $Ta=10^5$,  $\phi=\frac{\pi}{4}$ and $Q=100$ (thin solid line), $Q=500$ (dashed line), $Q=1000$ (dotted line), $Q=2000$ (squares), $Q=5000$ (thick solid line) and $Q=10000$ (dot-dashed line). Note $Q=2000$ and $Q=5000$ are omitted from (a) for clarity. $Q=100, 500$ and $1000$ are chaotic solutions, whereas $Q=2000, 5000$ and $10000$ are steady solutions.}\label{UVbar}
\end{center}
\end{figure}

To examine the vertical structure of \ubar and \vbar as a function of $z$, and its
dependence on $Q$, we plot the time-averaged mean flows in figure \ref{UVbar}. We expect
these plots to be more informative when considering \vbar than when considering $\bar u$,
as, from the time-dependent plots, we know that \ubar is highly fluctuating about zero,
however, we still examine both cases. All parameters are held constant and we explore a
range from $Q=100$ to $Q=10000$, each value of $Q$ is shown in a different line-type.
First, note how the size of \vbar changes as $Q$ is increased; from figure \ref{UVbar} (b) we see that with each increase of $Q$ comes a decrease in the maximum value of $\bar v$.  A slight change in the vertical structure of
\vbar is also evident. As $Q$ is increased from $100$ to $1000$, the layer depths at which
the maxima occur move towards the mid-layer depth, as we saw in figure \ref{tdmf}. From
$Q=2000$ to $10000$, the solutions are steady and perhaps should be considered separately,
though the $Q=2000$ and $Q=5000$ cases do have a similar vertical structure. However, the
$Q=10000$ case stands out, as the direction of $\bar v$ has reversed and its structure is
different. This will be examined in more detail in section \ref{sec:mfeqn}. 

As expected, the change in structure of \ubar is trickier to interpret as \ubar is more
time-dependent; we therefore omit the results for two values of $Q$ from the plot, for clarity (see figure \ref{UVbar} (a)). It is clear though, that as $Q$
is increased, the strength of the flow in the boundary layers is increased in the chaotic
regime. This contributes to the increase in $\langle \bar u \rangle_{rms}$ with $Q$ that was exhibited in figure \ref{mfvQ} (a). However, the behaviour of the bulk flow also contributes: as $Q$ is increased from $100$ to $500$, the flow in the bulk increases in size, but as $Q$ is increased to $1000$ the flow in the bulk is significantly suppressed. Hence the increase in boundary layer flow as the field strength in increased from $Q=500$ to $Q=1000$ is responsible for the corresponding increase in $\langle u \rangle_{rms}$. Whereas $\bar v$ maintained a consistent shear profile until $Q=10000$, $\bar u$ exhibits a number of structural changes as $Q$ is increased. For example, in the bulk, between $Q=100$ and $Q=500$, the flow has reversed direction. So increasing the magnetic field strength not only has an effect on the energy in the mean flows, it can also change the direction of the
mean flow. What causes the change in vertical structure of the flows we observe in figure
\ref{UVbar} will be examined in section \ref{sec:mfeqn}. Finally, by comparing the sizes
of \ubar and $\bar v$, in figure \ref{UVbar}, we see that \vbar is larger than $\bar u$ in
all cases. We comment that this is not physically realistic and is a shortcoming of our local model with periodic boundary conditions.

\subsection{Properties of the mean magnetic field}\label{mfield}
In addition to the mean flows we can consider the behaviour of the mean fields, $\bar B_x$
and $\bar B_y$, as $Q$ is increased. For small $Q$, magnetic field is expelled to the boundaries leaving the bulk of the layer with almost zero mean magnetic field. This behaviour is observed for both $\bar B_x$, which is initially zero, and $\bar B_y$, which is the direction of the imposed field. 
Flux expulsion has been reported in similar systems to the one we study here (see e.g., \cite{Arter}, \cite{Taoetal1998}). As $Q$ is increased, magnetic field is expelled to the boundaries to a lesser extent and so the mean field is increasingly abundant in the bulk of the layer.
We note that the behaviour changes for solutions in the steady regime (large $Q$) but as before we choose to focus on the behaviour of the chaotic regime.
The consequences of magnetic field being expelled to
the boundaries arise from the fact that, if there is little, or no, magnetic field in the
bulk of the layer, it will be unable to affect the mean flow there. We analyse the
competition between the flows and the fields in section \ref{sec:mfeqn}.

\subsection{Mean flow equations}\label{sec:mfeqn}
Equations governing the mean flows can be obtained by horizontally averaging the $x$ and $y$
components of the momentum equations, i.e.,
\begin{equation}\label{mfeq1}
 Pr\bar u = \frac{Pr}{Ta^{\frac{1}{2}}\sin\phi}\frac{\partial ^2 \bar v}{\partial
z^2}-\frac{1}{Ta^{\frac{1}{2}}\sin\phi}\frac{\partial \overline{vw}}{\partial z} +\frac{Pr
\zeta Q}{Ta^{\frac{1}{2}}\sin\phi}\frac{\partial \overline{B_yB_z}}{\partial z},
\end{equation}
\begin{equation}\label{mfeq2}
 Pr\bar v = -\frac{Pr}{Ta^{\frac{1}{2}}\sin\phi}\frac{\partial ^2 \bar u}{\partial
z^2}+\frac{1}{Ta^{\frac{1}{2}}\sin\phi}\frac{\partial \overline{uw}}{\partial z} -\frac{Pr
\zeta Q}{Ta^{\frac{1}{2}}\sin\phi}\frac{\partial \overline{B_xB_z}}{\partial z},
\end{equation}
where we have averaged in time and assumed a statistically steady state so that
$\frac{\partial }{\partial t}\langle \bar u \rangle=\frac{\partial }{\partial t}\langle
\bar v \rangle=0$.
We will refer to the term on the left-hand sides of the
equations as the Coriolis term, the first term on the right-hand sides as the viscous
term, the second term on the right-hand sides as the Reynolds stress (RS) term and the
last term on the right-hand sides as the Maxwell stress (MS) term. Notice that equations (\ref{mfeq1}) and (\ref{mfeq2}) have been normalised by a factor $Ta^{\frac{1}{2}}\sin\phi$; however, this is unimportant here as we will focus on the relative magnitudes of the different terms.
It is the balance between the terms of these equations that determines the size and structure of the mean flows. Note, it is the correlations of the
flow and the field in the $x$-direction with the flow and field in the $z$-direction that
dictate the mean flow in the $y$-direction. Similarly, it is the correlations of the flow and
the field in the $y$-direction with the flow and field in the $z$-direction that dictate
the mean flow in the $x$-direction. 

\begin{figure}
\begin{center}
\includegraphics[scale=0.76]{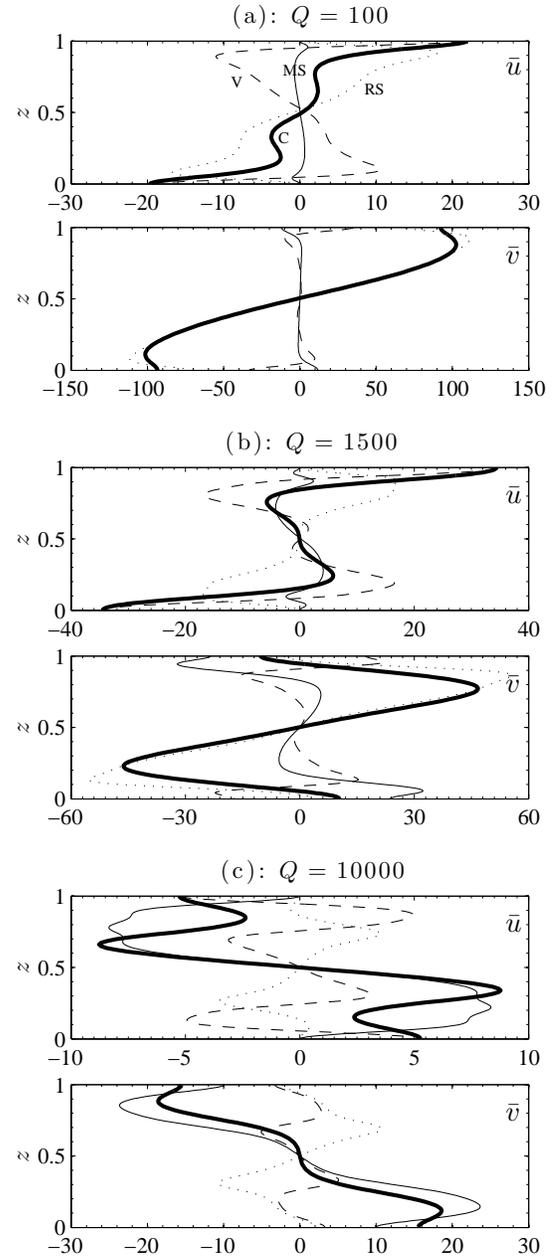}
\caption{Terms of the mean flow equations, (\ref{mfeq1}) (top panels) which drive $\bar u$, and
(\ref{mfeq2}) (bottom panels) which drive $\bar v$ for $Pr=1$, $\zeta=1.1$, $Ra=5\times10^5$, $Ta=10^5$,  $\phi=\frac{\pi}{4}$ and (a) $Q=100$, (b) $Q=1500$, (c) $Q=10000$. The Coriolis terms are given by the thick solid lines (these are equivalent to the mean flows themselves, as $Pr=1$), the RS terms are given by the dotted lines, the MS terms by the thin solid lines and the viscous terms by the dashed lines.}\label{dombal}
\end{center}
\end{figure}

In section \ref{mf}, we saw that increasing $Q$ had an effect on the size and
structure of the mean flows. To understand what is dictating this change, we plot each of
the terms of the mean flow equations, (\ref{mfeq1}) and (\ref{mfeq2}), as a function of
$z$. The plots are shown in figure \ref{dombal} for (a) $Q=100$, (b) $Q=1500$ and (c)
$Q=10000$.

First, let us consider the case when the field strength is small, $Q=100$. In the bottom
plot of figure \ref{dombal} (a), we see clearly that the dominant balance is between \vbar and the RS term, with the MS and viscous
terms making only a small contribution. We see that the extrema of the RS terms are close
to the boundaries resulting in a mean flow with maximum value close to the boundaries.
Similarly, from the top plot of (a), \ubar is driven by the RS term. However,
in this case there is a larger contribution from the viscous term, a likely result from the fact
that \vbar is greater than $\bar u$. The MS term, when $Q=100$, is small compared to the other terms. For small $Q$,
the MS term is expected to be small for two reasons: firstly (and most obviously), the MS term is proportional
to $Q$ and secondly, as we discussed in section \ref{mfield}, for $Q=100$,
$\bar B_y$ is particularly small in the bulk owing to flux expulsion, suggesting that the correlations $\overline{B_yB_z}$ are
likely to be small in the bulk too. Since the MS term is small, we are left with a similar
balance as in the hydrodynamic case.

Increasing $Q$ to $Q=1500$ gives the balance shown in case (b). \vbar is clearly still
driven by the RS term. However, the field has acted to reduce the magnitude of the RS term and as a result, the mean flow is also suppressed. As $Q$ has increased, the MS terms have become larger. In particular, the MS term driving $\bar v$ is most
significant close to the top and bottom boundaries, and since it is acting in the opposite
direction to the RS term there, it reduces the size of the mean flow driven and so \vbar is
relatively small in these boundary layer regions compared to \vbar in the bulk of the
layer. At this $Q$, the magnetic field is strongest near the boundaries and because the magnitude of $Q$ is large enough, the MS term is
significant at the boundaries, resulting in the behaviour we observe. Some field does exist in
the bulk though and so the MS terms have started to have an effect there too.
Furthermore, the increase in the effect of the MS term, along with the fact that the
maximum of the RS term has moved towards the middle of the layer (compared with smaller
$Q$), mean that the maximum of \vbar has also also moved towards the mid-layer depth. This
explains the behaviour of $\bar v$ observed in the time-dependent plots of figure
\ref{tdmf}.
For $\bar u$, the increase in $Q$ to $Q=1500$ has similarly resulted in an increase in the MS term affecting it. However, the RS term is of comparable size to the RS term when $Q=100$. The RS, MS and viscous terms are all significant in determining $\bar u$, they act such that the direction of $\langle \bar u \rangle$ is reversed in the bulk (compared to $Q=100$) and so that it is slightly larger in size. At the boundaries, there are relatively large viscous boundary layers and
since close to the boundaries the RS and MS terms are small, it follows that \ubar has
boundary layers where the flow is largest, in agreement with the plots in figure
\ref{tdmf} and \ref{UVbar}. 

Increasing $Q$ further, to $Q=10000$ (see figure \ref{dombal} (c)), leads to the MS terms
becoming the dominant terms. Magnetic field is no longer expelled to the boundaries; this
fact combined with the large $Q$ means that the MS terms are dominant across the whole
layer. The RS terms modify the flows but they have been sufficiently suppressed that it is the MS term that
dominates the structure. For this reason, \vbar is in the opposite direction for
$Q=10000$ than it is for the other $Q$ shown in figure \ref{UVbar}. 

Hence, we have shown that the field can act to change the direction of the flow through
changing which terms in equations (\ref{mfeq1}) and (\ref{mfeq2}) are dominant. In the
cases examined, increasing $Q$ does not appear to change the direction of the mean flow by
changing the direction of the Reynolds stresses; instead, it does it through suppressing the Reynolds stresses so that they are no longer the dominant term.

\section{Effect of decreasing the Prandtl numbers}\label{results2}
In this section, we decrease $Pm$ whilst also ensuring $Pr<1$ and $\zeta<1$. This is towards a regime more appropriate for stellar interiors. We consider $Pr<\zeta<1$ so that $Pm<1$. In particular, we take $Pr=0.1$ and $\zeta=0.5$ corresponding to $Pm=0.2$, this is in contrast to $Pm=0.91$, the value throughout section \ref{results1}.  Hereafter, we refer to the parameters used in section \ref{results1} as case (i) and those in this section as case (ii). In order to maintain approximately the same degree of rotational constraint as in case (i) (as measured by $Ro$) and also approximately the same degree of nonlinearity at $Q=0$, we set $Ra=2.5\times10^5$ and $Ta=5\times10^5$. The critical Rayleigh numbers associated with case (ii) are also shown in figure \ref{Critical}.

\subsection{Effect on the flow regime}
\begin{figure*}
\begin{center}
\includegraphics[scale=1]{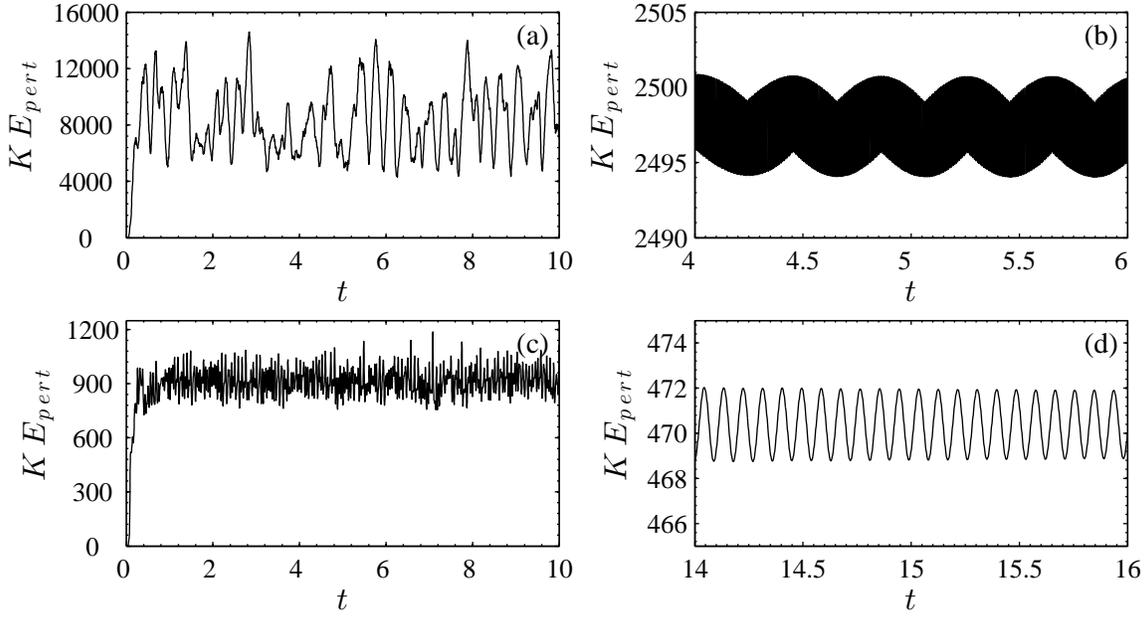}
\caption{Different types of solution encountered as $Q$ is increased for $Pr=0.1$, $\zeta=0.5$, $Ra=2.5\times10^5$, $Ta=5 \times10^5$ and $\phi=\frac{\pi}{4}$. In (a), $Q=300$ and solution is of bursting-type, in (b) $Q=2000$ and the solution is quasi-periodic, in (c), $Q=4000$ and the solution is chaotic and in (d) $Q=6200$ and the solution is periodic.}\label{KEvt}
\end{center}
\end{figure*}

At these parameters, as the strength of the imposed magnetic field is increased (through increasing $Q$), a number of different types of solution are encountered. For $Q$ less than approximately $1250$, the convection exhibits signs of bursting (described in more detail below). Increasing $Q$ further leads to small region of quasi-periodic (QP) behaviour before becoming chaotic (without bursts) at $Q\sim2000$. The solutions then remain chaotic until large $Q$, at which point the solutions becomes periodic and eventually steady. To illustrate the differences in behaviour in each of these regimes, we include a time series of $KE_{\text{pert}}$ for examples in the different regimes (see figure \ref{KEvt}).

\begin{figure}
\begin{center}
\includegraphics[scale=0.95]{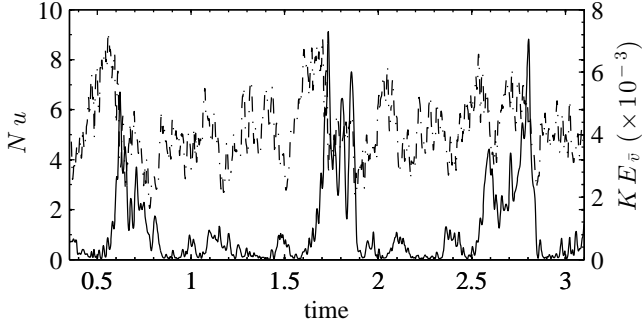}
\caption{Behaviour of the Nusselt number, $Nu$ (dashed line) and the kinetic energy in $\bar v$, $KE_{\bar v}$, (solid line) for a bursting-type solution. The large peaks in $Nu$ come just before the peaks in $KE_{\bar v}$.}\label{bursting}
\end{center}
\end{figure}

The bursting solutions were not evident in the simulations of section \ref{results1} and so we describe them in more detail here. As the convection gets more vigorous, a larger mean flow is driven, this flow then acts to inhibit the convection and in doing so kills its own source of energy. Therefore the reduction in convection is closely followed by a reduction in the strength of the mean flow. Once the mean flow is quenched the convection can build up again and the process repeats, each cycle resulting in the bursts of energy we see in fig \ref{KEvt}(a). An illustration of this is seen in figure \ref{bursting} where a peak in the Nusselt number, $Nu$ (a measure of convective efficiency) directly precedes a peak in the kinetic energy of $\bar v$, $KE_{\bar v}$. Other bursting solutions have been seen in studies of convection in different systems (e.g., \cite{BrummellHart1993, GroteBusse2001, RotvigJones2006, Teedetal2012}).

\begin{figure*}
\begin{center}
\includegraphics[scale=1]{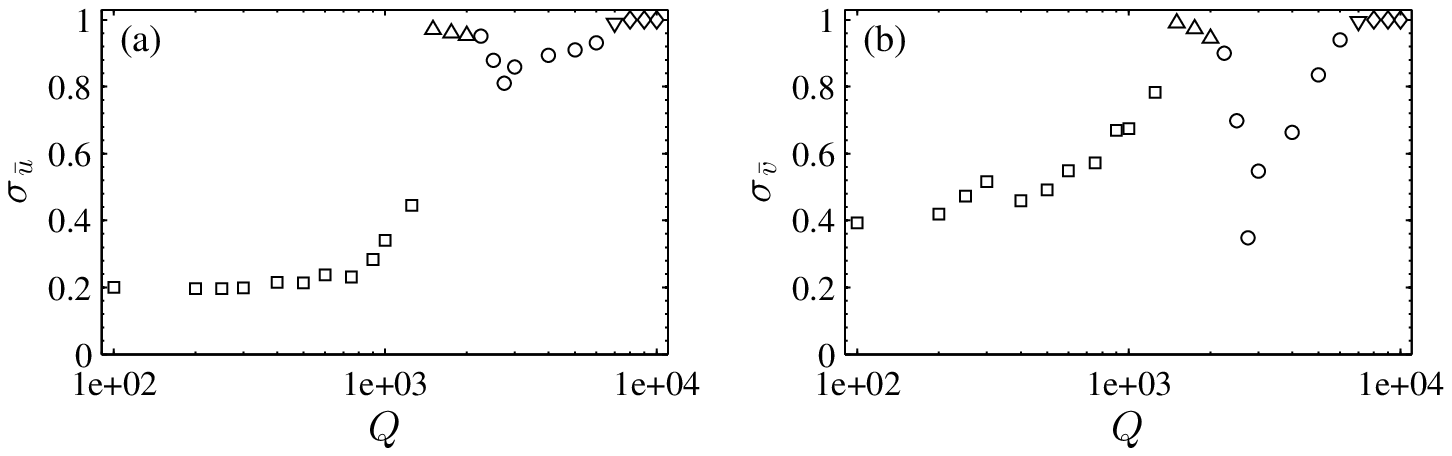}
\caption{(a) $\sigma_{\bar u}$ and (b) $\sigma_{\bar v}$ for $Pr=0.1$, $\zeta=0.5$, $Ra=2.5\times 10^5$, $Ta=5\times 10^5$ and $\phi=\frac{\pi}{4}$. Bursting-type solutions are marked with a square, quasi-periodic with an upwards facing triangle, chaotic with a circle, periodic with a downwards facing triangle and steady with a diamond.}\label{systematic2}
\end{center}
\end{figure*}

In figure \ref{systematic2} we show $\sigma_{\bar \xi}$ as function of $Q$ for $\bar u$ and $\bar v$, marking each different type of solution with a different symbol. As in figure \ref{systematic}, we see that increasing the strength of magnetic field increases $\sigma_{\bar \xi}$, i.e., the fluctuations in time become fewer. Comparing with figure \ref{systematic}, we see that the behaviour of $\sigma_{\bar u}$ is very similar for both $Pm$ but that $\sigma_{\bar v}$ is significantly smaller, especially at small $Q$, in the small $Pm$ case. This difference comes around because the bursting type regime results in a much larger variation from the mean when averaging than the original chaotic solutions. We find that in this case, the bursts are more pronounced in the energy of $\bar v$ compared to $\bar u$.

\subsection{Effect on mean flow generation}
\begin{figure*}
\begin{center}
\includegraphics[scale=1]{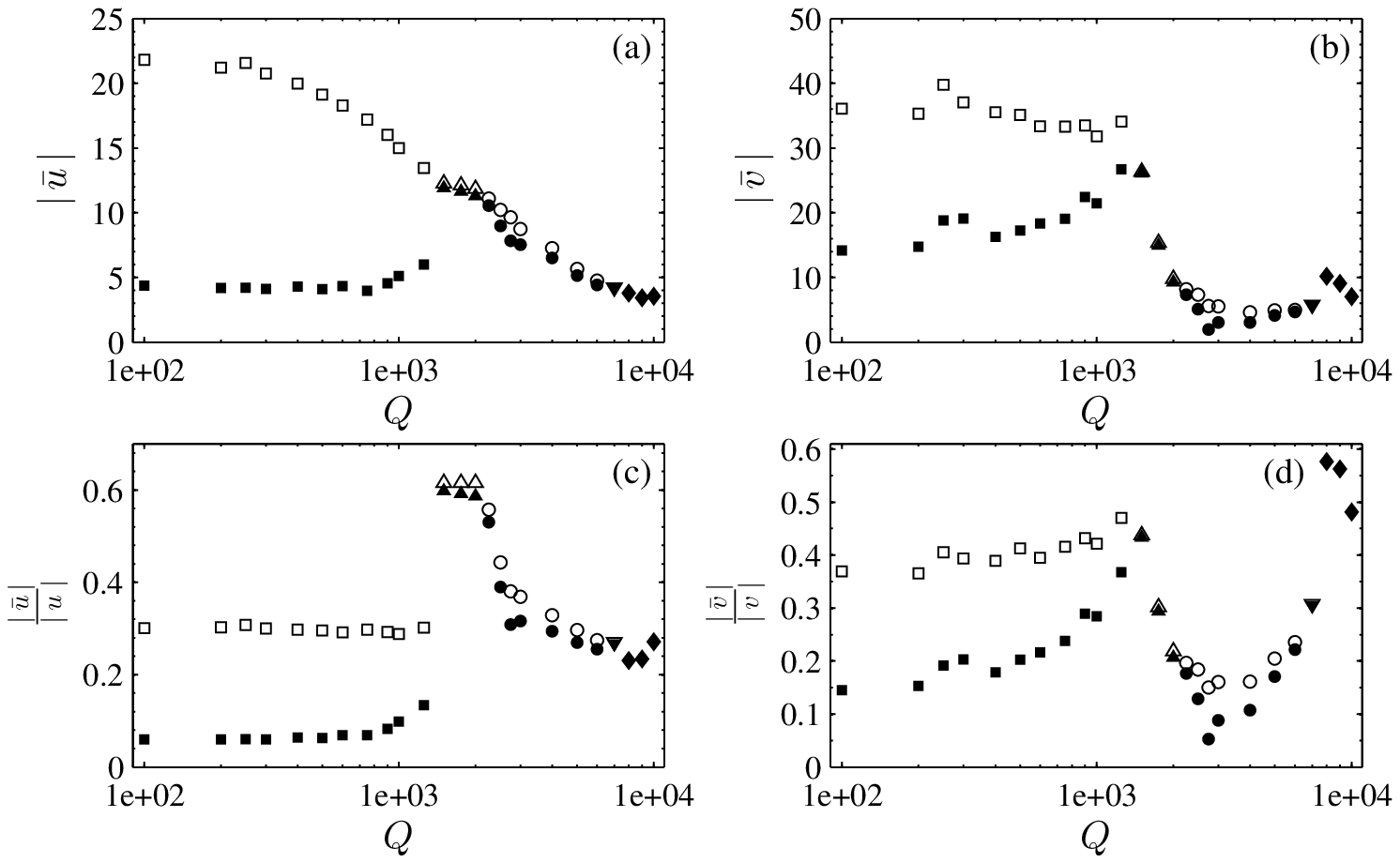}
\caption{Size of $\bar u$ and $\bar v$ for the same parameters as in figure \ref{systematic2}. (a) shows the size of $\bar u$ as given by (\ref{one}) (closed symbols) and (\ref{two}) (open symbols). (b) shows the equivalent for $\bar v$. (c) gives the ratio of the size of $\bar u$ to the size of $u$ for the size of $\bar u$ given by (\ref{one}) (closed symbols) and by (\ref{two}) (open symbols). (d) shows the equivalent ratios for $\bar v$. The shape of the symbol has the same meaning as in figure \ref{systematic2}.}\label{mfvQ2}
\end{center}
\end{figure*}

We consider whether small $Pr$ and $\zeta$ inhibit or promote mean flow generation and maintenance in the presence of a horizontal magnetic field. We analyse the size of $\bar u$ and $\bar v$ as given by the two measures in (\ref{one}) and (\ref{two}) (see figure \ref{mfvQ2} (a) and (b)). It is clear each of the different regimes exhibits different behaviour and so we consider each in turn. 
The bursting-type solution is characterised by the difference in the two measures; $\bar \xi_{r\langle m \rangle s}$ is large compared to $\langle \bar \xi \rangle _{rms}$. In this regime, $\bar u _{r\langle m \rangle s}$ decreases significantly as $Q$ is increased, whereas $\bar v_{r\langle m \rangle s}$ maintains a fairly constant value. On the other hand, $\langle \bar \xi \rangle _{rms}$ increases for both $\bar u$ and $\bar v$ as $Q$ is increased.
Increasing $Q$ further leads to a short quasi-periodic regime, here both measures of $\bar u$ and $\bar v$ are decreasing, albeit $\bar v$ is doing so more severely than $\bar u$. For $\bar u$ this trend continues as $Q$ is further increased and the solutions become chaotic and then periodic. Whereas, for $\bar v$, the size of the mean flow decreases but then increases at the largest $Q$. 

As we did for case (i), we consider the effect of increasing $Q$ on the ratio of the size of $\bar \xi$ to the size of $\xi$, (see figure \ref{mfvQ2} (c) and (d)). For $\bar u$, $\bar u_{r \langle m \rangle s}/|u|$ remains roughly constant in the bursting- type regime, suggesting that the decrease in $\bar u_{r\langle m \rangle s}$ as $Q$ increases results from the decrease in $u$ itself. However, since $\langle \bar u \rangle_{rms}$ increases with $Q$, the ratio does too. Now, for $\bar v$, in the bursting-type regime, $\bar v_{r \langle m \rangle s}$ remains roughly constant with increasing $Q$ but since $|v|$ decreases with increasing $Q$, the ratio , $\bar v_{r \langle m \rangle s}/|v|$ is increasing with $Q$. That is, the magnetic field suppresses the velocity field more than it does the correlations driving $\bar v$. As expected, the increase in $\langle \bar v \rangle _{rms}$ with $Q$ results in an increase in the ratio too.
For the non-bursting solution regimes, the ratios of $\bar u_{r \langle m \rangle s}$ and $\langle \bar u \rangle_{rms}$ to $|u|$ tend to decrease with $Q$ and so there the mean flows are suppressed by the field more than the flow itself. For $\bar v$, this behaviour is seen until approximately $Q=2500$ (where the size of $\bar v$ starts to increase again) and then the ratio (as it has to) increases with $Q$.

Comparing these results with those in case (i) (cf. figure \ref{mfvQ}) we find that the size of the mean flows in case (i) are larger. However, for $\bar u$ the ratios are comparable in the chaotic/bursting-type regimes (i.e., at small to moderate $Q$) whilst for $\bar v$, the ratios are larger in case (i). This difference arises from the bursts that are clearly evident in $\bar v$ for case (ii) but not case (i). The biggest difference between cases (i) and (ii) is evident in the solutions for $Q\lesssim1250$ (i.e., where case (i) is chaotic and case (ii) exhibits signs of bursting). In case (i), $\langle \bar v \rangle_{rms}$ decreases with increasing $Q$, whereas in cases (ii) it increases. Also, $\bar v_{r \langle m \rangle s}$ is steeply decreasing in case (i) as opposed to remaining roughly constant in case (ii). This difference results in a situation where in case (i) the field suppresses the mean more than the velocity field itself but in case (ii), the mean flow is suppressed by a smaller amount (if at all) than the flow itself. Therefore, for $\bar v$, it appears that in the smaller $Pr$, $\zeta$ case, the field has a smaller effect on the mean flow. In other words, the magnetic field switches off the velocity perturbations in both cases (i) and (ii) but in the case of small $Pr$ and $\zeta$, the field in not as efficient at switching off the mean flow.

\subsection{Balances of the mean flow equations}
\begin{figure}
\begin{center}
\includegraphics[scale=0.76]{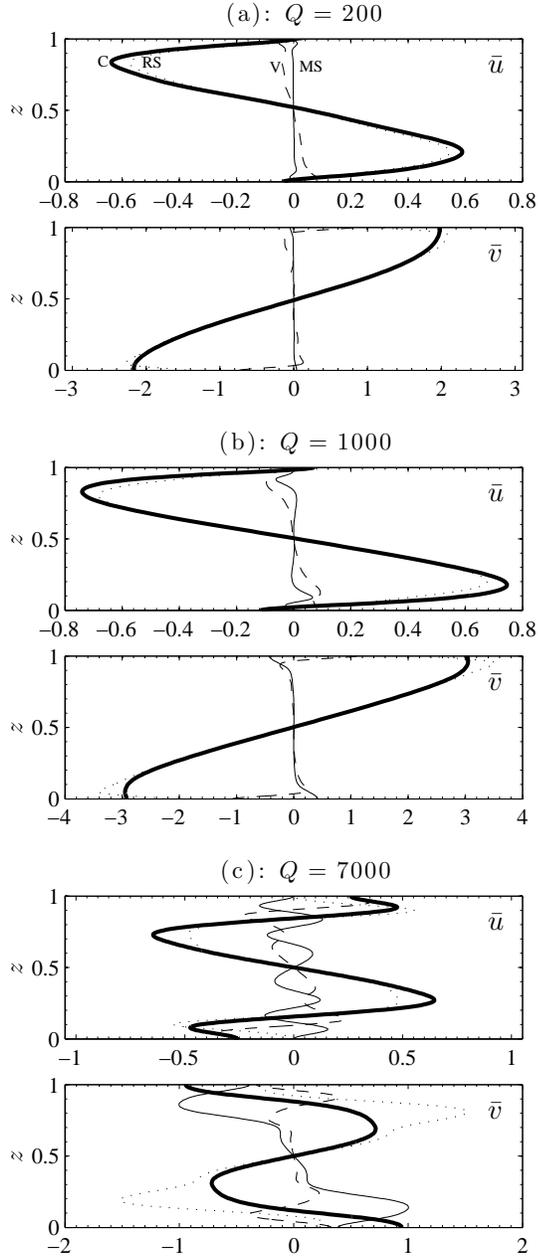}
\caption{Terms of the mean flow equations, (\ref{mfeq1}) (top panels) which drive $\bar u$, and
(\ref{mfeq2}) (bottom panels) which drive $\bar v$ for $Pr=0.1$, $\zeta=0.5$, $Ra=2.5\times10^5$, $Ta=5\times10^5$, $\phi=\frac{\pi}{4}$ and (a) $Q=200$, (b) $Q=4000$, (c) $Q=7000$. The Coriolis terms are given by the thick solid lines, the RS terms are given by the dotted lines, the MS terms by the thin solid lines and the viscous terms by the dashed lines. The mean flows themselves (not shown) are a scale factor of $10$ larger than the Coriolis terms.}\label{dombal2}
\end{center}
\end{figure}

As we saw in section \ref{sec:mfeqn}, it is a balance between the RS, MS and viscous terms that determines the size and vertical structure of $\bar u$ and $\bar v$; which of these terms (or combination of terms) dominated depended upon the parameters considered. Here we assess how the behaviour of these terms changes for different $Q$ at small $Pr$ and $\zeta$. Immediately, we expect the viscous term to be much less important as it is proportional to $Pr$ which we have decreased by a factor of $10$ (this is confirmed to be the case in figure \ref{dombal2}). Therefore, the mean flows are determined by a direct competition between the RS and MS terms. The decrease in $\bar u_{r\langle m\rangle s}$ as $Q$ is increased (seen in figure \ref{mfvQ2} (a)) results from a decrease in the time-dependent RS term as well as an increase in the time-dependent MS term, whilst the increase in $\langle \bar u \rangle_{rms}$ with $Q$ arises because of an increased coherence in the correlations in the RS term such that the time-averaged RS term increases sufficiently to overcome the increase in the time-averaged MS term and cause an increase in the size of $\bar u$. A similar argument can be made to explain the increase in $\langle \bar v \rangle_{rms}$ with $Q$.

In figure \ref{dombal2} we display the terms of equations (\ref{mfeq1}) and (\ref{mfeq2}) for three values of $Q$ in case (ii). Again, it is clear that (as expected for small $Pr$), the viscous term plays a much less significant role than it did in driving the flows in case (i). In (a) and (b), the terms correspond to flows that are in the bursting-type regime and we see that in both cases the flows are dominated by the RS terms; even at $Q=1000$, the MS terms are not large enough to have an impact, again this is likely to be due to the $Pr\zeta$ factor in the MS terms.
It is also clear that, as explained above, the time-dependent RS term increases in size as $Q$ in increased form $200$ to $1000$, and this results in the larger $\bar u$ and $\bar v$ at $Q=1000$. When $Q$ is increased to $7000$ (c), the RS terms are suppressed significantly and the MS term has a larger impact, these effects combine to suppress the mean flows at this $Q$. Even though the MS terms are now important, the vertical structure of the flows is still dominated in the bulk by the RS terms.
However, increasing $Q$ further eventually leads to a regime in which the MS terms do dictate the mean flows driven, as they did in figure \ref{dombal} (c).

Further comparisons of the plots in figure \ref{dombal2} with the equivalent ones for case (i) (figure \ref{dombal}) show a striking difference to be the reversal in the direction of $\langle \bar u \rangle$ at small $Q$; it is clear this reversal is a result of a reversal in the direction of the RS term driving $\bar u$. 

In summary, for the parameters in case (ii), the RS terms are not suppressed in the way they were in case (i) and the MS term is less significant for $Pr=0.1$, $\zeta=0.5$, this allows the mean flows to persist when $Q$ is increased even with a decrease in the overall fluid velocity.

\section{Discussion and conclusions}\label{conc}
The main aim of this paper was to examine the effect of a horizontal magnetic field on mean flow generation by rotating convection in two dimensions. In general, the field acts to suppress the fluid velocity but we have showed it has more complicated interactions with the processes that drive mean flows. We focussed on two sets of examples: one at $Pr=1$ and $\zeta=1.1$ (case (i)) and one at $Pr=0.1$ and $\zeta=0.5$ (case (ii)) whilst approximately maintaining the rotational constraint through fixing $Ro$. 

In both cases (i) and (ii), at small to moderate $Q$ (i.e., small to moderate magnetic field strengths), we illustrated that the field acts to reduce the average magnitude of the time-dependent, horizontally-averaged flows but that the field also organises these flows so that they are more systematic in time. We also demonstrated the effect of decreasing $Pr$ and $\zeta$. In case (i), increasing the imposed field strength affects the processes driving $\bar v$ as much (if not more) than it did $v$ itself, but in case (ii), where $Pr$ and $\zeta$ were decreased, the magnetic field suppresses $v$ much more than it did $\bar v$. In other words, the magnetic field appears to be less effective at suppressing the mean flow (as opposed to the overall flow) when $Pr$ and $\zeta$ are small. However, this change in behaviour is also accompanied by a difference in the type of solution observed at small to moderate $Q$ in cases (i) and (ii). In case (i), the solutions are chaotic, whereas in case (ii), the solutions, whilst chaotic, also exhibit bursting tendencies. Further investigation is required to establish if it is solely the small $Pr$ and $\zeta$ that result in $\bar v$ being able to persist as the imposed field strength is increased, or if the change to a bursting regime that coincided with the decrease in $Pr$ and $\zeta$ is responsible. Any clear trends in the differences in the effect of the imposed magnetic field on $\bar u$ were harder to establish.

By analysing the horizontally-averaged (mean) equations we revealed what was responsible for the size and vertical structure of the mean flows at different $Q$. In general, a balance between the MS and RS terms drives the flows, though at the relatively modest values of $Ra$ and $Ta$ considered here, the viscous term plays a role when $Pr=1$. 
At small $Q$, magnetic flux is expelled to the boundaries causing the field to have a significant effect on the behaviour close to the boundaries but almost no effect on the bulk fluid. These effects lead to the size and structure of the bulk mean flows being dominated by the RS terms and, in some cases, modified by the viscous terms. However, in case (i), as $Q$ is increased, the RS terms are suppressed by the field and the MS terms have a larger impact; these two processes act together to suppress the mean flows in this regime. 

In case (ii), the smaller $Pr$ means that the viscous terms are much less important. Furthermore, the small $Pr$ and $\zeta$ result in the RS terms dominating the flows at larger $Q$, as the MS terms do not contribute as they do in case (i). Indeed, a much larger $Q$ has to be reached in the smaller Prandtl number regime (case (ii)) than is required in case (i) for the MS terms to dominate. In case (ii), at small to moderate $Q$, the time-averaged RS terms actually increase with field strength and so this, coupled with the fact the MS terms contributed less, leads to mean flows that are able to persist even though the velocity field is being suppressed by the field.

Our results emphasise that the interaction between mean flows and magnetic fields is quite complex and, in particular, they highlight the crucial role of the Prandtl numbers. Our results show strong dependence on these parameters: in case (i) it is clear that magnetic field influences the balance between the Reynolds and Maxwell stresses; however, how robust this behaviour is is still unclear as shown by the results at smaller Prandtl numbers (case (ii)). Whilst our model is a crude simplification of the full problem, the demonstration that large-scale flows may be able to persist in the presence of a magnetic field has potential applications to astrophysical flows (e.g., differential rotation in stars).

To conclude, we recognise the limitations of our crude model. Firstly, in two dimensions, correlations may be amplified resulting in strong mean flows and flow suppression. We are therefore currently examining how extending the model to three dimensions affects the driving of mean flows and their suppression by magnetic fields. Furthermore, the periodic boundary conditions we imposed are likely to be unrealistically enhancing the meridional flows relative to the zonal flows. 
The modest parameters used in this work were for illustrative purposes and are orders of magnitude below astrophysically relevant values ($Ra$, $Ta> 10^{10}$). However, with modern-day computing facilities, there is the potential to probe more realistic regimes; this forms an avenue for prospective follow-up work.
Finally we note that, throughout this study, we imposed a uniform horizontal magnetic field. In reality, magnetic fields are generated and sustained by dynamo action; an obvious problem to address then is whether the mean flows generated in our system are capable of sustaining a magnetic field through dynamo action, this is something we address in a future paper.

\acknowledgments
Most of this work was carried out whilst I was a Ph.D student at the University of Leeds and I am grateful to Science and Technology Facilities Council (STFC) for a Ph.D. studentship. This work was also supported by the European Research Council under ERC grant agreements No. 337705 (CHASM). Some of this work was undertaken on ARC1, part of the High Performance Computing facilities at the University of Leeds, UK, and some on the University of Exeter supercomputer, a DiRAC Facility jointly funded by STFC, the Large Facilities Capital Fund of BIS and the University of Exeter. I would also like to thank Steve Tobias and Matthew Browning for many helpful discussions whilst carrying out this work, as well as an anonymous referee for useful comments that improved the paper.

\listofchanges

\end{document}